\documentclass[useAMS,usenatbib,usedcolumn]{mn2e}

\def\HI{\ifmmode{\rm HI}\else{H\/{\sc i}}\fi}

\def\lsun{\ifmmode{{\mathrm L}_{\odot}}\else{L$_{\odot}$}\fi}

\def\msun{\ifmmode{{\mathrm M}_{\odot}}\else{M$_{\odot}$}\fi} 
\def\msunpc2{\ifmmode{{\mathrm M}_{\odot} \, {\mathrm{pc}}^{-2}}\else{M$_{\odot} \, {\mathrm {pc}}^{-2}$}\fi}

\def\kms{\ifmmode{{\mathrm{km \, s^{-1}}}}\else{${\mathrm{km \, s^{-1}}}$}\fi}

\def\la{\mathrel{\mathchoice {\vcenter{\offinterlineskip\halign{\hfil
$\displaystyle##$\hfil\cr<\cr\sim\cr}}}
{\vcenter{\offinterlineskip\halign{\hfil$\textstyle##$\hfil\cr
<\cr\sim\cr}}}
{\vcenter{\offinterlineskip\halign{\hfil$\scriptstyle##$\hfil\cr
<\cr\sim\cr}}}
{\vcenter{\offinterlineskip\halign{\hfil$\scriptscriptstyle##$\hfil\cr
<\cr\sim\cr}}}}}
\def\ga{\mathrel{\mathchoice {\vcenter{\offinterlineskip\halign{\hfil
$\displaystyle##$\hfil\cr>\cr\sim\cr}}}
{\vcenter{\offinterlineskip\halign{\hfil$\textstyle##$\hfil\cr
>\cr\sim\cr}}}
{\vcenter{\offinterlineskip\halign{\hfil$\scriptstyle##$\hfil\cr
>\cr\sim\cr}}}
{\vcenter{\offinterlineskip\halign{\hfil$\scriptscriptstyle##$\hfil\cr
>\cr\sim\cr}}}}}

\def\aj{AJ}
\def\araa{ARA\&A}
\def\apj{ApJ}
\def\apjl{ApJ}
\def\apjs{ApJS}
\def\aap{A\&A}
\def\aaps{A\&AS}
\def\mnras{MNRAS}
\def\pasp{PASP}

\usepackage[figuresright]{rotating}
\usepackage{lscape}
\usepackage{graphics}
\usepackage{epsfig}
\usepackage{amsmath}
\usepackage{amssymb}
\usepackage{lscape}

\title[Ionized gas discs in E/S0s at $z<1$]{Ionized gas discs in elliptical and S0 galaxies at  $z<1$}
\author[Y.~Jaff\'e et al.] {Yara L. Jaff\'e$^{1}$\thanks{E-mail: yara.jaffe@astro-udec.cl}, Alfonso Arag\'on-Salamanca$^2$, Bodo Ziegler$^3$,  Harald Kuntschner$^4$, 
\and  Dennis Zaritsky$^5$, Gregory Rudnick$^6$,  Bianca M. Poggianti$^7$,  Carlos Hoyos$^8$, \and Claire Halliday$^9$, Ricardo Demarco$^1$.\\ 
   $^1$Department of Astronomy, Universidad de Concepci\'on, Casilla 160-C, Concepci\'on, Chile\\
   $^2$School of Physics and Astronomy, The University of Nottingham, University Park, Nottingham NG7 2RD, UK \\
   $^3$Department of Astrophysics, University of Vienna, T\"{u}rkenschanzstr. 17, 1180 Wien, Austria\\
   $^4$European Southern Observatory, Karl-Schwarzchild Strasse 2, D-85748 Garching, Germany\\
   $^5$University of Arizona, 933 N. Cherry Ave, Tucson, AZ 85721, USA \\  
   $^6$Department of Physics and Astronomy, The University of Kansas,  Malott room 1082, 1251 Wescoe Hall Drive, Lawrence, KS, 66045, USA\\ 
   $^7$INAF-Osservatorio Astronomico di Padova, Vicolo dell Osservatorio, 5, I-35122, Padova, Italy\\
   $^8$Departamento de Astronomia, Instituto de Astronomia, Geof\'sica e Ci\^encias Atmosf\'ericas da Universidade de S\~ao Paulo,  Rua do Mat\~ao 1226, \\
   Cidade Universit\'aria, 05508-090, S\~ao Paulo, Brazil\\
   $^9$23, rue d'Yerres, F-91230 Montgeron, France\\   
}

\begin{document}

\date{Submitted: 21 November 2013. Accepted: 12 March 2014}

\maketitle

\begin{abstract}
We analyse the extended, ionized-gas emission of 24 early-type galaxies (ETGs) at $0<z<1$ from the ESO Distant Cluster Survey (EDisCS). 
We discuss different possible sources of ionization and favour star-formation as the main cause of the observed emission. 
10 galaxies have disturbed gas kinematics, while 14 have rotating gas discs. In addition, 15 galaxies are in the field, while 9 are in the infall regions of clusters. This implies that, if the gas has an internal origin, this is likely stripped as the galaxies get closer to the cluster centre. 
If the gas instead comes from an external source, then our results suggest that this is more likely acquired outside the cluster environment, where galaxy-galaxy interactions more commonly take place. 
We analyse the Tully-Fisher relation of the ETGs with gas discs, and compare them to EDisCS spirals. Taking a matched range of redshifts, $M_{B}<-20$, and excluding  galaxies with large velocity uncertainties, we find that, at fixed rotational velocity, ETGs are 1.7 mag fainter in $M_{B}$ than spirals. 
At fixed stellar mass, we also find that ETGs have systematically lower specific star-formation rates than spirals. 
This study constitutes an important step forward towards the understanding of the evolution of the complex ISM in ETGs by significantly extending the look-back-time baseline explored so far. 
\end{abstract}


\section{Introduction}
\label{sec:introduction}

On the basis of their observed properties, early-type galaxies (ETGs) have long been regarded as a homogeneous population of passively-evolving galaxies. For example, ETGs follow a tight correlation between colour and magnitude \citep[][]{Baum1959,VS1977} that distinguishes them from the (blue) population of spiral and irregular galaxies. The small scatter about this relation suggests a predominantly old stellar population \citep{BLE,Aragon1993,Jaffe2011}. Furthermore, 
ETGs present a striking correlation between their effective radius, mean surface brightness, and velocity central dispersion \citep[the Fundamental Plane; e.g.][]{Faber1987,Dressler1987,DjorgovskiDavis1987}. Combined evidence from studies of the Faber-Jackson, Mgb-$\sigma$, Fundamental Plane, and line strengths, provide support for  the passive evolution of cluster ETGs \citep[see e.g.][and references therein]{Ziegler2001,Fritz2009}. 
Finally, studies of the chemical composition and alpha enhancement in galaxies have shown that more massive galaxies tend to be more metal rich and have a shorter star formation time-scale \citep{Thomas2005}.

However, the idea that ETGs are red and dead systems essentially devoid of gas and dust, has been questioned in the last decades. Owing to an increase in instrumental sensitivity, a number of observations have gradually revealed the presence of a complex intra-stellar medium (ISM) in ETGs. Some examples include the detection of hot gas through X-rays \citep[][]{Forman1979,OSullivan2001,Macchetto96}, HI gas \citep[e.g.][]{VerdesMontenegro2001,GopalKrishna2012}, warm ionized gas \citep[][]{DemoulinUlrich1984,Kim1989,Trinchieri1991,Kehrig2012}, and dust \citep{Rowlands2012,diSerego2013}. 
In this context, \citet{Trager2000} proposed a `frosting' model in which the apparently young ages inferred for some ellipticals by single stellar population models, is due to a ``frosting'' of younger stars in a primarily old stellar population \citep[but see][]{MarastonThomas2000}. 
More recently, significant advances  have been made in the study of ETGs by the SAURON and ATLAS$^{\rm 3D}$ collaboration using integral-field spectroscopy \citep[see][for a recent overview]{Cappellari2011b}. Their studies of local E/S0s confirmed that emission is commonly found in the central regions of elliptical galaxies and even more so in S0s \citep[e.g.][]{Phillips86,Macchetto96,Sarzi2006,Yan2006}, and also revealed a great level of complexity in the internal kinematics of these galaxies. 
Their most robust explanation for the gas found in some of these galaxies is that it has been accreted from an external source, owing to the coexistence of co- and counter-rotating components of both the gas and the stars in many cases \citep[][]{Shapiro2010}.

In spite of all these efforts, the origin and state of the gas in ETGs remains a puzzle. 
Many studies have been devoted to understanding the source of the gas ionization in ETGs  \citep[][]{Binette1994,stasinska2008,Sarzi2010,Annibali2010,Finkelman2010,Kehrig2012}. 
Different excitation mechanisms have been proposed, including photo-ionization of the inter-stellar medium (ISM) by either post asymptotic giant branch (AGB) stars, shocks, active galactic nuclei (AGN), or OB stars. It  remains unclear, however, the degree to which these processes contribute to the ISM ionization of ETGs 

One way to understand the formation and evolution of ETGs is by studying their scaling relations \citep[e.g. the Fundamental Plane, Faber-Jackson relation, or colour-magnitude relation, see][for a review]{Renzini2006}
 as a function of redshift, environment and intrinsic galaxy properties \citep[e.g.][]{vanderWel2004,Jorgensen2006,Saglia2010,Jaffe2011}. To further understand the evolutionary link between the different galaxy types, it is also useful to investigate the location of ETGs in the late-type galaxy scaling relations. 
For disc galaxies, an especially useful relation is the Tully-Fisher relation \citep*[TFR;][]{TullyFisher1977}, which relates the disc luminosity to the maximum rotational velocity of disc galaxies. 
Studies of the TFR as a function of morphology have shown significant differences between E/S0 and spiral galaxies. 
 \citet{Gerhard2001} studied the TFR of nearly round elliptical galaxies and found that they follow a TFR with marginally shallower slope than spiral galaxies and $\sim1\,$mag fainter zero-point in the $B$-band.   
\citet{Bedregal2006} also found an offset in the TFR of S0s and spirals.  Naively, this offset can be interpreted as the result of the fading of spiral galaxies since they ceased forming stars. The large intrinsic scatter however suggests that the S0s cannot have simply faded after having (all) transformed at a single epoch.  They conclude that the scatter in the S0 TFR arises from the different times at which galaxies began their transformation. 
More recently, \citet{Williams2010} performed a careful study of the S0 TFR also finding an offset between the spiral and the S0 TFR. They extensively analyse biases and obtain a smaller (but still significant) TFR offset than in \citet{Bedregal2006}.  They conclude from their study that this offset can be explained if S0s are systematically smaller or more concentrated than spirals. 
However, the recent study of \citet{Rawle2013} of Coma spirals and S0s essentially confirms  the conclusions and interpretation of \citet{Bedregal2006}. Moreover, they find that the amount of fading experienced by the S0s is correlated not only with the time since the cessation of star formation, but also with the time of accretion into the cluster, clearly indicating that the transformation of spirals to S0s is accelerated by cluster environment.

All of the above-mentioned studies however have been done at low redshift. In this paper we present, for the first time, observational evidence for extended emission discs in ETGs at intermediate redshift ($0 < z < 1$) and perform a comparison with similarly selected emission-line spirals.

In \citet[][]{Jaffe2011b}, we studied the gas kinematics (from emission-lines), morphological disturbances, and the TFR of distant galaxies as a function of environment using the ESO Distant Cluster Survey (EDisCS). In our analysis, the vast majority of the emission-line galaxies were spirals and irregulars. However, we found that a significant number of ETGs at $0<z<1$ have extended emission in their spectra. 
Such high-redshift objects, which have not yet been studied,   
are the primary focus of this paper. In Section~\ref{sec:data} we summarize the EDisCS data and the parent sample of \citet[][]{Jaffe2011b}. We present the emission-line ETG sample in Section~\ref{sec:el_etg_sample}, along with their gas dynamics and stellar morphologies.  
In Section~\ref{sec:results}, we explore the possible mechanisms responsible for ionizing the gas in the emission-line ETGs, and then study their environments, TFR, and specific star formation rates (SFRs).  We then discuss our results in Section~\ref{sec:discusion} and summarize our main findings and conclusions in Section~\ref{sec:conclusions}.

\section{The parent sample}
\label{sec:data}

The galaxies used in the analysis of this paper are drawn from the sample used in \citet{Jaffe2011b}, which consists of emission-line galaxies from the ESO Distant Cluster Survey (EDisCS). We summarize the EDisCS dataset briefly in Section~\ref{subsec:ediscs} and the sub-sample of \citet{Jaffe2011b} in Section~\ref{subsec:el_sample}, but refer to the original papers for a full description.

\subsection{EDISCS}
\label{subsec:ediscs}

EDisCS is a multi-wavelength survey of 20 fields containing galaxy clusters at redshifts between 0.4 and 1. The cluster sample was selected from 30 of the highest surface-brightness candidates in the Las Campanas Distant Cluster Survey \citep{Gonzalez2001}.

The dataset includes  optical photometry \citep[see][]{White2005} from FORS2 on the Very Large Telescope (VLT) in B, V, and I-band for the 10 intermediate-redshift cluster candidates and V, R, and I-bands for the 10 high redshift cluster candidates.
In addition, near-IR J and K photometry was obtained for most clusters using SOFI at the New Technology Telescope (NTT;  Arag\'on-Salamanca et al., in preparation). 
We also compiled deep multi-slit spectroscopy with FORS2/VLT \citep{Halliday2004,MJ2008}, which consists of high signal-to-noise data for $\sim 30 - 50$ members per cluster and a comparable number of field galaxies in each field down to $I = 22$ and  $I=23$ for the mid- and high-redshift clusters respectively. The wavelength range was typically $\sim$5300-8000 $\rm \AA{}$ for two of the runs and 5120-8450 $\rm \AA{}$ for the other two. 
Cluster and field galaxies were separated using spectroscopic redshifts \citep[see][for details]{MJ2008}.

In addition to this, for 10 of the higher redshift clusters from the database we acquired Hubble Space Telescope (HST) mosaic images in the F814W filter with the Advanced Camera for Surveys Wide Field Camera \citep[see][for details]{Desai2007}.

\subsection{The emission-line galaxy sample of \citet{Jaffe2011b}}
\label{subsec:el_sample}

To construct a TFR, in \citet{Jaffe2011b} we used a sub-sample of the EDisCS dataset consisting of galaxies with measurable emission in their spectra (typically the [OII]3727$\rm \AA$ doublet, H$\beta$, the [OIII]4959 and 5007$\rm \AA{}$ lines, H$\gamma$, and/or  H$\delta$). Given the 2D nature of the spectra, this selection was carried out by the careful visual inspection of each emission line. Subsequent quantitative analysis of the S/N of these lines revealed that in the majority of cases the fraction of pixels whose signal was larger than $2\sigma$ above the noise was over 95\% of all the detected pixels (i.e., those pixels accepted by the fitting procedure).
To ensure that rotation could be measured, we rejected galaxies with inclinations of less than $30^{\circ}$ (inclination $= 0$ corresponding to face-on). We also rejected observations affected by slit misalignment (misalignment with respect to the major axis of the galaxy  $>30^{\circ}$). After applying these selection criteria, we built a sample of 422 galaxies with 1024 emission lines in total.

The ``true'' parent emission-line galaxy distribution is well represented by this sample. 
The fraction of EDisCS galaxies with  emission-line spectra for which we were able to model emission lines and measure a rotation curve is fairly constant ($\simeq 35$\%) across the magnitude range of our galaxies.

In \citet{Desai2007}, galaxy morphologies were assigned by the visual inspection of HST images by a team of expert classifiers. The HST data however only covers about half of the full spectroscopic sample. For the HST sub-sample, we found that the morphology distribution of the emission-line galaxies was dominated by spirals and irregulars \citep[see Fig. 8 in][]{Jaffe2011b}, although we  identified 44 galaxies with E/S0 morphology.

\section{The emission-line ETG sample of this paper}
\label{sec:el_etg_sample}

In \citet{Jaffe2011b}, we focused mainly on the visually classified spiral galaxies within the emission-line galaxy sample of Section~\ref{subsec:el_sample}. This paper focuses now on the ETGs.  
We apply the selection used in \citet[][see also Section~\ref{subsec:el_sample}]{Jaffe2011b} to create the emission-line sample, and using the visual morphologies of \citet{Desai2007}, we identify 44 ETG candidates with emission lines (27 E and 17 S0). 
To ensure that the morphologies were reliable, three co-authors (YJ, AAS, and CHo) carefully re-examined the HST images and found that 3 galaxies were incorrectly classified as E/S0. Taking these misclassifications out of the sample, we were left with 41 ETGs, which is the sample considered in this paper. Table~\ref{all_table} shows the main properties of these galaxies, including the 3 misclassified ones.

\subsection{Emission-line kinematics}
\label{subsec:kinematics}

To make sure that the emission in the ETG sample is real, and to study the gas kinematics, we carefully inspected the two-dimensional (2D) emission-lines of each galaxy spectra, as well as the residuals resulting from subtracting 2D fits to each line, in the same manner as in \citet[][]{Jaffe2011b}. 
When checking the spectra, we found that only 24 of the 41 ETGs have real emission. 
The remaining 17 galaxies present ``artificial emission'' that is likely the result of artefacts in the 2D spectra and were thus discarded 
\citep[e.g. a poorly subtracted overlapping sky line or cosmic rays; see discussion in ][]{Jaffe2011b}. 
We further inspected the kinematics of the galaxies with real emission and found that 14 have undisturbed (rotating) disc kinematics, and 10 have clear asymmetries in their emission-line profiles (i.e. disturbed kinematics), as shown in Figures~\ref{HST_mosaic_ETG_rot}~and~\ref{HST_mosaic_ETG_dist} respectively. As explained in \citet{Jaffe2011b}, rotation velocities could only be safely computed for those galaxies with undisturbed disc kinematics. 
For those galaxies however, the emission was typically quite extended \citep[$\sim$2 to 5 times the photometric scalelength for S0s and ellipticals respectively, see fig.~21 in][]{Jaffe2011b}, so a reliable rotation velocity could be measured. In terms of the effective radius ($r_{eff}$), the emission in these ETGs typically extends up to $1-5 \times r_{eff}$ (see columns 9~and~10 of Table~\ref{all_table}).

Unfortunately, with the available data, we cannot measure the velocity dispersion (of either the gas or the stars) for our sample.

\subsection{The final emission-line ETG sample}
\label{subsec:finalsample}

\begin{figure}
\begin{center}
  \includegraphics[width=0.49\textwidth]{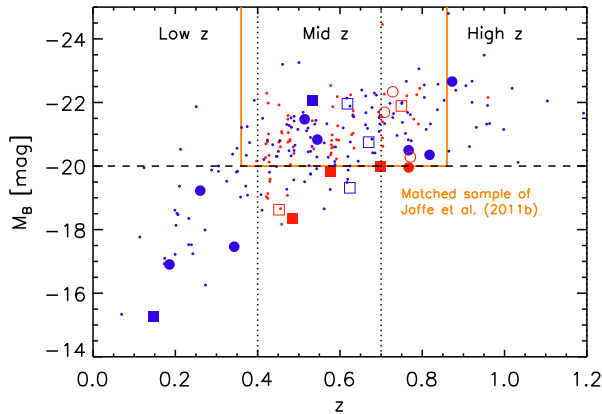}
\end{center} 
\caption{Redshift vs. $M_{B}$ for the parent emission-line sample with spiral morphology (small dots), and the ETG sample (larger symbols). In both cases, blue symbols correspond to field galaxies and red ones to galaxies in clusters or groups. For the ETG sample,  we have distinguished ellipticals (circles) from S0s (squares), as well as galaxies with disturbed (open symbols) and undisturbed kinematics (filled symbols). The solid orange box corresponds to the matched sample used in \citet[][sample ``C" in their paper]{Jaffe2011b}.  The vertical dashed lines indicate the different redshift ranges used in the analysis of this paper (e.g. Figs~\ref{TFR}~and~\ref{SFR_mass}), whilst the horizontal dashed line places a reasonable $M_B$ limit to compare the galaxies across all redshifts.}
 \label{matched_sample}
\end{figure}

After the morphological and dynamical inspections, the final true emission-line ETG sample consists of 24 galaxies, that is 14 ellipticals and 10 S0s, among which 14 have undisturbed and 10 disturbed kinematics. Their main properties are listed in the top part of Table~\ref{all_table}.  
Note that 2 of the galaxies in the final sample are a close pair (EDISCS ID's have an appended ``\_A'' and ''\_B'') and hence,  suffer from mutual light contamination. 
For this reason, although we include these galaxies in our sample, we do not analyse their individual kinematics.

Figures~\ref{HST_mosaic_ETG_rot} and~\ref{HST_mosaic_ETG_dist} show the HST postage-stamp images of all the emission-line ETGs in our sample, as well as their single-Sersic model and residuals (cf. Section~\ref{subsec:morph}), 2D spectra and emission-line model.

As Figure~\ref{matched_sample} shows, our sample spans a broad range of redshifts and absolute rest-frame $B$-band magnitudes ($M_{B}$, corrected for Galactic extinction).
Values of $M_{B}$ were calculated from the observed spectral energy distribution of each galaxy, normalized to its total $I$-band flux, and the spectroscopic redshift \citep[see][for details]{Rudnick2009}.  
The distribution of ETGs in this plot is very similar to that of the spirals. 
In \citet{Jaffe2011b}, we defined a ``matched sample´´ of cluster and field galaxies by imposing the following magnitude and redshift cuts:  $M_B < -20$ and $0.35 < z < 0.86$ (orange square in the figure). 
Owing to the broader redshift range of the ETGs considered in this paper and the small number of galaxies, we define three redshift bins (low, intermediate, and high) to pursue our comparative analysis (c.f. Section~\ref{sec:results}). These are separated by vertical dotted lines in  Figure~\ref{matched_sample}. We indicate with a horizontal dashed line the magnitude limit up to which we can reliably compare galaxies at all redshifts ($M_{B}=-20$). This line was chosen to roughly match the completeness of the sample across all the redshifts considered.

We emphasize that, within the matched sample of \citet[][yellow box in Figure~\ref{matched_sample}]{Jaffe2011b}, ETGs with emission lines are quite rare: only $\sim 12$\% of the emission-line galaxies have early-type morphology. Conversely,  about $18$\% 
of E/S0s in the matched sample have detected emission lines. 
It is likely, however, that ETGs with either weak emission lines or slit misalignments were rejected from our selection (more so than spirals). Hence, the true fraction of ETGs with gas discs is  likely larger than our estimate.

\section{Results}
\label{sec:results}

We first explore the possible sources of ionization of the ETGs presented in Section~\ref{subsec:finalsample}, and then  
study their environments, morphological disturbances, TFR,  and specific SFRs (sSFRs).

\begin{figure}
\begin{center}
  \includegraphics[width=0.49\textwidth]{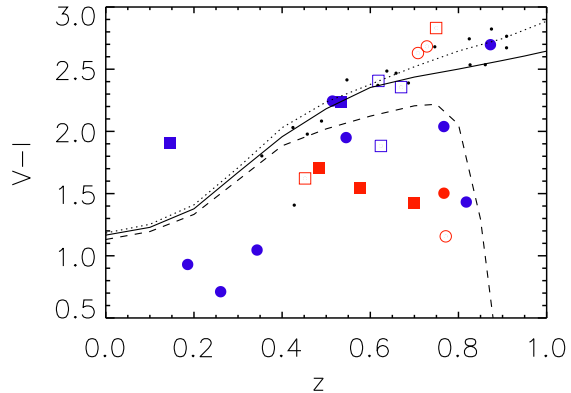}
\end{center} 
\caption{We compare the $V-I$ observer frame colour and \ redshift data  for both the  cluster (red) and field (blue) ETGs  with the evolutionary synthesis models of \citet[][lines]{bc03}. As in Figure~\ref{matched_sample}, ellipticals are represented as circles, S0s as squares, kinematically disturbed galaxies as open symbols, and undisturbed ones as filled symbols. Note that no magnitude or redshift cuts were applied to the plotted sample. 
The black dashed line corresponds to a ``formation redshift" (resdhift of last star-formation episode) of $z_{\rm F}\sim0.9$, the solid line to $z_{\rm F}\sim1.5$, 
and the dotted line to $z_{\rm F}\sim2.5$, as in \citet[][Fig.~12]{Jaffe2011}. All models assume solar  metallicity ($Z_{\rm solar}=0.02$). For reference, the smaller black points correspond to the ETGs with ``artificial emission'' (i.e., galaxies whose apparent emission is an observational artefact, see section~\ref{subsec:kinematics}).}
 \label{col_z}
\end{figure}

\begin{figure}
\begin{center}
  \includegraphics[width=0.5\textwidth]{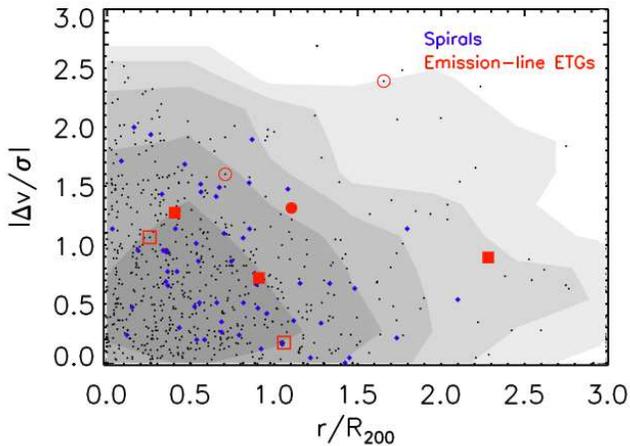}
\end{center} 
\caption{Stacked phase-space diagram ($r$/$R_{200}$ vs. $|\Delta v/\sigma|$)  for all the EDisCS cluster and group galaxies (grey points, no magnitude or redshift cuts were applied). The grey contours trace the number density distribution of the cluster galaxies, that concentrate, as expected, at low $r$/$R_{200}$ and $|\Delta v/\sigma|$. The emission-line galaxy sample with visual morphologies discussed in this paper is highlighted with larger coloured points. The smaller blue diamonds are the spiral population of emission-line galaxies. As in Figure~\ref{matched_sample}, the red circles and squares are elliptical and S0 galaxies, respectively. For the E/S0s, filled symbols trace galaxies with regular (rotating) emission and open circles disturbed kinematics. 
Interestingly, none of the cluster E/S0s in our sample are located in the virialized part of the cluster (bottom-left corner of the plot). Instead, they reside in the cluster infall region. Note that the galaxy EDCSNJ1054525-1244189 is not plotted, as it belongs to a very low mass group (182 km s$^{-1}$ of velocity dispersion) with only 10 members, and hence its phase-space position is not physically meaningful. }  
 \label{r_200}
\end{figure}

\subsection{Source of ionization}
\label{subsec:ionization}

We have considered three different possible causes for the ionization of the gas in ETGs, namely i)  AGN activity, ii) star formation, or iii) post-AGB stars. Although the available FORS2 spectra are not ideal to unambiguously test the possible scenarios, we discuss each case below and place some constraints.

In \citet{Poggianti2006} it was estimated that the contamination from pure AGN in the EDisCS spectroscopic sample is at most 7\%.  Although this is a statistical result, it is unlikely that all of our ETGs belong to this 7\% minority. Owing to the limited spectral coverage of most of our spectra, we are unable to identify AGN using a BPT diagram \citep{baldwin81}. However, following \citet{SanchezBlazquez2009}, we investigated, when possible, the ratio of [OII] to $H_{\beta}$. For the 2 ETGs with detected [OII] and $H_{\beta}$  we find that [OII]/$H_{\beta}$ is much lower than $6.7$, the limit below which the emission is considered to be powered by star formation \citep[based on the condition $\lceil {\rm OII} \rceil$/${\rm H}_{\alpha} < 1.5$, from][]{Kewley2004}. Moreover, AGN are expected to exhibit centrally-concentrated nuclear emission, 
contrary to what we observe in our ETG sample (see  Figures~\ref{HST_mosaic_ETG_rot}~and~\ref{HST_mosaic_ETG_dist}, and Table~\ref{all_table}). Notwithstanding the limited available evidence, we are inclined to favour star-formation over AGN activity. 

The other possibility is that the ionization of the gas in ETGs is mainly caused by hot evolved stars rather than young ones \citep{Trinchieri1991}. This scenario has been largely debated in recent years \citep[see e.g.][and references therein]{Boselli2005,Yi2011,Zaritsky2014} and has received strong observational support from various studies \citep[][and references therein]{YanBlanton2012}. In particular, \citet{Binette1994} showed that, over a wide range of ionization parameteres, photoionization by post-AGB stars can result in an [OII]/H$_{\alpha}$ ratio below 2.5, which translates to [OII]/H$_{\beta} <7$. Our measurements of  [OII]/H$_{\beta}$  alone thus cannot rule out ionization from an evolved stellar population.

We further investigated the locations of our emission-line ETGs in the colour-magnitude diagram and compared them with those of the other EDisCS ETGs. Following the same approach used in  \citet{Jaffe2011}, we used colours that were the nearest to rest-frame $U-V$ ($R-I$ for most clusters, and $V-I$ for the lowest redhift clusters), and found that some of the cluster ETGs  displaying emission (kinematically disturbed and undisturbed) have colours that are consistent with the red sequence, but most are somewhat bluer. 

For the field galaxies, we used the evolutionary synthesis models of \citet{bc03} to predict the observed $V-I$ colours of galaxies at different redshifts, assuming a passively evolving stellar population that formed in a single burst of 0.1Gyr duration, as in \citet[][]{Jaffe2011}. 
This is shown in Figure~\ref{col_z} for different models, together with the observed $V-I$ colours of the field (blue symbols) and cluster (red) emission-line ETGs, that are also  plotted for comparison. The different lines correspond to models computed with different times since the last star-formation episode, $t_{\rm F}$. Such look-back times can also be expressed in terms of a ``formation redshift", $z_{\rm F}$ \citep[see][for details]{Jaffe2011}. 

In Figure~\ref{col_z}, the emission-line ETGs have colours that are, on average, 0.3 magnitudes bluer than the passive model lines.  More precisely, almost half of these ETGs (47\%) have colours bluer than this average, and the colours can be up to 1 magnitude bluer than the models. 
Such colours can be explained by a small amount of ongoing or recent star formation, involving as little as $\sim 5$\% of the galaxy stellar mass, happening in the past $\sim 1\,$Gyr. The exact amount of recent star formation required, naturally depends on the detailed star-formation history of the galaxies and the time of observation. Our estimate however is reasonable for a broad range of star-formation histories \citep[see e.g.][]{Barger1996}. 

In addition to colours, we looked at the  Balmer absorption  ($H_{\delta}$ and $H_{\gamma}$) strength and the 4000$\rm \AA$ break ($D_{n,4000}$) of the galaxies, as they are sensitive to the galaxies' star-formation histories. They are also less susceptible to reddening than rest-frame optical colours.  Measures of these quantities are described in  Rudnick et al. (in preparation), but we note here that the line strengths were measured by decomposing the emission and absorption following \citet{Moustakas2010}. 
In their paper, Rudnick et al. examine the [OII] emission of galaxies with old stellar populations, as defined by their EW($H_{\delta}$+$H_{\gamma}$)/2 and $D_{n}(4000)$. They distinguish  ``young", ``intermediate", and ``old" stellar populations based on the luminosity-weighted stellar age sequence seen in the  EW($H_{\delta}$+$H_{\gamma}$)/2--$D_{n}(4000)$ plane \citep[see e.g.][]{Kauffmann2003}, and through the comparison with stellar evolution  synthesis models. 
They  conclude from their study that the ionized gas observed likely stems from a combination of mass loss and accretion, and is stripped by ram-pressure in clusters. 

When placing our emission-line ETGs in the $D_{n}(4000)$ vs. EW($H_{\delta}$+$H_{\gamma}$)/2  diagram of Rudnick et al., we find that   most (69\%) of the emission-line ETGs are consistent with having young-to-intermediate stellar populations, whilst only 4 (out of 13) galaxies are consistent with having ``old" stellar populations. Note that, given that most of our galaxies have evidence of significant recent star formation, they would not all have been included in Rudnick et al.'s sample. Nevertheless, more than half (13 out of 24) of our ETGs could be used in the comparison, and the characteristics of this subsample trace well the entire population of emission-line ETGs.

In sum, we conclude that star formation is the most likely explanation for the emission observed in most of the ETGs considered, although we cannot completely rule out the contribution of post-AGB stars or AGN to the ionization of all (or some) of the galaxies.

\subsection{Environment}
\label{subsec:environment}

Across the full redshift and $M_{B}$ range of EDisCS galaxies, the emission-line ETG sample is dominated by field galaxies (62\% of the final sample, blue symbols in Figure~\ref{matched_sample}). Only  9 emission ETGs were classified as  cluster or group members (38\%, red symbols). This is partly because there are no EDiSCS clusters at either $z\lesssim0.35$ or at the highest redshifts, which leads to field galaxies outnumbering cluster galaxies in the EDisCS spectroscopic sample.  However, the difference is small: 55\% of EDisCS galaxies are in the field, while 45\% are in clusters or groups. Moreover, if we take the matched sample region of \citet[][orange box in Figure~\ref{matched_sample}]{Jaffe2011b}, which was defined to compare cluster and field galaxies, we find that emission-line ETGs are still predominantly field galaxies (70\%). 

The field population is dominated by kinematically undisturbed galaxies, with only a third of  field ETGs having distorted gas kinematics. In contrast, many of the cluster ETGs are kinematically disturbed galaxies, and also are all located towards the cluster outskirts, as shown in Figure~\ref{r_200}. 
The plot displays the projected distance to the cluster centre, $r/R_{200}$,  on the x-axis, and the absolute value of the peculiar velocity, $|\Delta v/ \sigma_{cl}|$, on the y-axis, where $\Delta v =v_{gal}-v_{cl}$, $v_{gal}$ is the line-of-sight velocity of the galaxy, $v_{cl}$ the central velocity of the cluster, and $\sigma_{cl}$ the cluster velocity dispersion.  The grey points are all the cluster galaxies in EDisCS \citep[defined as being within $\pm $ $3\sigma _{\rm cl}$ from $z_{\rm cl}$; ][]{MJ2008}, the blue ones are the cluster spiral galaxies within the emission-line sample of \citep{Jaffe2011b} and the red large symbols correspond to the cluster E/S0s discussed in this paper. 
As shown by the work of \citet[][]{Mahajan2011,Oman2013}, and references therein, the location of galaxies in projected phase-space can give an idea of the time that the galaxies have spent in the cluster: within the bottom-left corner region in the plot, galaxies are mainly virialized, whilst the intermediate and outer regions (at higher velocities and/or distance to the cluster centre) contain a higher fraction of backsplash and infalling galaxies.  
The location of the few cluster ETGs with emission in phase-space indicates that they probably fell very recently into the cluster environment. 

Our findings clearly indicate that ETGs with emission are preferably in the field, or in the outskirts of galaxy clusters.

\subsection{Morphological Disturbances}
\label{subsec:morph}

The ETGs considered in this paper are all elliptical and S0 galaxies with no strong signs of interaction or morphological distortions in their HST images.  However, to identify the presence of small morphological disturbances undetected in our visual examination of the HST images,  we visually inspected the residuals of single-S\'ersic profile fits to the images, using the method described in  \citet[][]{Hoyos2011}.  This method is ideal to find interaction relics, as it enhances the residuals by subtracting the bulk of the smooth symmetric light. 
After this procedure, we identified 8 ETGs with disturbed morphologies (20\% of emission-line ETGs, see column 9 of Table~\ref{all_table}), 5 of which inhabit the field. Half of them have disturbed kinematics and the other half have regular disc kinematics; this is consistent with \citet{Jaffe2011b}, where we found, using the entire emission-line galaxy sample, that disturbances in the gas distribution and kinematics are unrelated to  morphological (stellar) disturbances. 
We emphasize that the disturbances discussed here are subtle and they do not represent major merger events or interactions. 

The upper panels of Figures~\ref{HST_mosaic_ETG_rot} and~\ref{HST_mosaic_ETG_dist} show the morphologies and single-Sersic fits for the final ETG sample separated by their kinematic properties (see Section~\ref{subsec:kinematics}).

\begin{figure*}
\begin{center}
\includegraphics[width=0.99\textwidth]{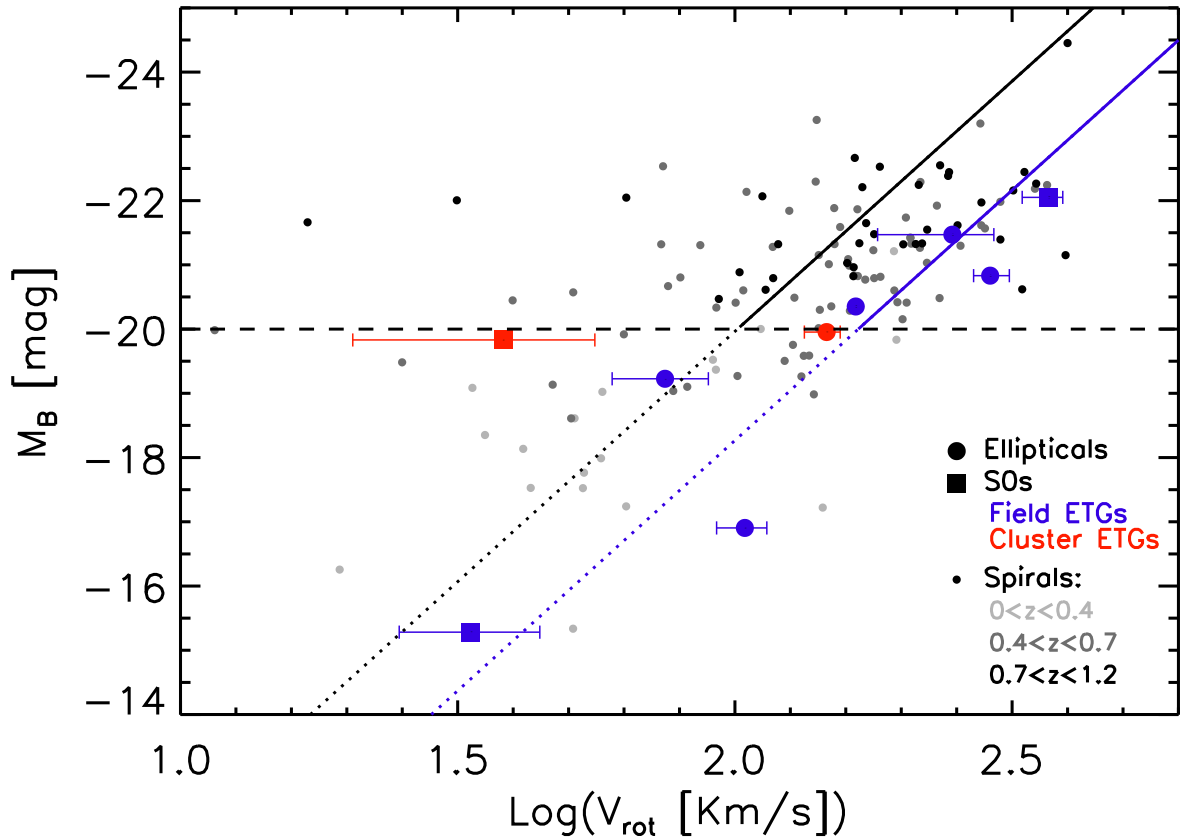}
\end{center} 
\caption{We compare the absolute-magnitude $M_B$ and rotation velocity data of EDisCS spirals (small symbols) and ETGs (bigger symbols). The spiral data are colour-coded in terms of redshift, as indicated in the bottom-right corner of the plot. Ellipticals (large solid circles) and lenticulars (large filled squares) are also distinguished. A red colour indicates that a galaxy is in a cluster/group environment, while blue corresponds to the field. We do not make redshift or magnitude cuts to the plotted sample, although we do not plot galaxies with velocities and velocity uncertainties consistent with no rotation (See galaxies labelled with ${^\dagger}$ in  Table~\ref{all_table}). The solid lines correspond to the fitted TFR assuming the slope of \citet{Tully1998} and considering either the median shift in $M_{B}$ for the spiral galaxies brighter than  $M_{B}=-20$ (black solid line), or the ETGs brighter than the same value (blue solid line). This value represents the magnitude limit above which below which the completeness is constant across the redshift range considered. For reference, we also show the relations extrapolated to the fainter galaxy magnitudes (dotted lines). 
}
 \label{TFR}
\end{figure*}

\subsection{The Tully-Fisher relation}
\label{subsec:TFR}

We present the TFR of emission-line ETGs and compare them to the spiral relation. We use the kinematically undisturbed galaxies, as in \citet{Jaffe2011b}, where we found that  global environment does not affect the TFR.  
To check that these galaxies had non-zero rotation, we further rejected galaxies with $V_{rot} < 2\sigma^{-}_{V_{rot}}$, where $\sigma^{-}_{V_{rot}}$ is the 95\% level uncertainty in the rotational velocity, $V_{rot}$ \citep[see][]{Jaffe2011b}. Five of our 14 kinematically undisturbed ETGs were consistent with no rotation and thus excluded from the TFR. The proportion of rejected galaxies does not depend on morphology, so applying this condition does not bias our results.

To construct the emission-line TFR, we derived galaxy inclinations by fitting a bulge and a disc to F814W HST images\footnote{The two-component fit performed to derive inclinations is independent of the single-Sersic fit made to the HST images to examine possible interaction relics (Sec.~\ref{subsec:morph}). }.
These two-component fits assumed that the bulge has a de Vaucouleurs profile and the disc an exponential profile, both convolved to the PSF of the images. This was done using the GIM2D software \citep[see][for a detailed description of the method used]{Simard2002,Simard2009}.  
Inclinations were used to correct absolute magnitudes for internal extinction \citep[see][]{Tully1998} and compute rotation velocities.

We emphasize that presence of a ``disc'' component does not necessarily imply that there is an actual stellar disc, because many dynamically hot systems have simple exponential profiles. We also note that there might be slight biases in the inclination corrections ($sin(i)$) applied to the emission-line rotational velocities because the galaxy inclinations were measured for the galaxy stellar-light distributions. This could cause an overestimation of the rotation velocity of ellipticals that, unfortunately, cannot be accounted for. Moreover, the gaseous disc may not have a similar stellar counterpart. 

Figure~\ref{TFR} shows the $B$-band TFR of the spiral galaxies in the parent sample (small dots), as well as that of ETGs (larger symbols).  
Although we plot all the galaxies without any redshift or magnitude cut, to make a fair comparison between the spiral and ETG samples we only consider galaxies with $M_{B}<-20$ and we colour-code the spiral galaxies by redshift (see Figure~\ref{matched_sample} for justification). 

Despite the large scatter, it is clear that, for the  $M_{B}$-limited sample, the emission-line ETGs are fainter than their spiral counterparts at a fixed rotational velocity. 
Applying the same slope found by \citet[][]{Tully1998} for local galaxies, we fit the TFR by computing the median shift in $M_{B}$ (for galaxies with $M_{B}\leq-20$), for the EDisCS spiral (solid black line) and ETG (solid blue line) populations separately. 
The dotted lines are only extensions of the bright TFR to fainter magnitudes for reference. 
The difference between the spiral and ETG $B$-band TFRs is 1.7 magnitudes in the bright regime. 

The rest-frame $B$-band is sensitive to both stellar population age and recent/current star formation, which could indicate that, if the emission in ETGs is caused by star formation (see discussion in Section~\ref{subsec:ionization}), this is significantly lower than that found in spirals.

It has been shown that the scatter about galaxy scaling relations can be reduced by analysing the ratio between the rotational velocity and the galaxy velocity dispersion \citep[$V_{rot}/\sigma$, e.g.][]{Kassin2007,Zaritsky2012}. 
Utilizing this quantity,  \citet{Zaritsky2012} illustrated how all S0s (as well as ellipticals as $V_{rot}/\sigma$ increases) deviate away from the fundamental plane. 
This is also true for the TFR, where they find that S0s can be shifted to the spiral relation when adding the pressure term.  
As mentioned in Section~\ref{subsec:kinematics}, we cannot retrieve  information about the velocity dispersion from our spectra. We note however, that the dependence on the pressure term is only important when studying stellar kinematics. In this paper, we do not consider stellar dynamics but instead use emission lines (gas) to study rotation.

\subsection{Specific star formation rates}
\label{sec:ssfr}

If we assume that star formation is the main cause of the emission in the ETGs considered in this paper (see discussion in Section~\ref{subsec:ionization}), we can investigate their star-formation properties by using [OII]3727$\rm \AA$  emission as a star-formation proxy. We cannot use other estimators such as H$_{\alpha}$, which is not in our spectral coverage, or infrared (IR) data, because the available far IR images only detect very high levels of star formation \citep[$\gtrsim 10.3 M_{\odot} yr^{-1}$ at $z = 0.6$;][]{Vulcani2010} and none of the ETGs of this paper were detected.

We use the SFRs, uncorrected for dust, presented in \citet{Poggianti2008} when available. 
These fluxes were obtained by multiplying the observed [OII] equivalent width by the continuum flux, estimated from the broadband photometry using total galaxy magnitudes. We further compute specific SFRs (sSFRs) by normalizing the SFR by stellar mass \citep[as in][]{Vulcani2010}. We note that the sSFRs computed here are upper limits, as some (if not most) of the ionization could come from post-AGB stars or AGN, as discussed in Section~\ref{subsec:ionization}. 

\begin{figure}
\begin{center}
  \includegraphics[width=0.5\textwidth]{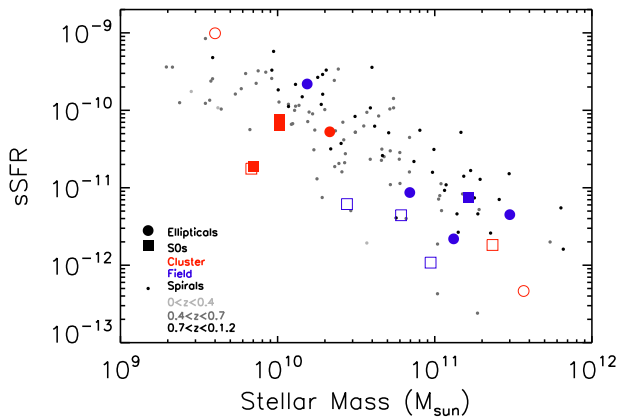}
\end{center} 
\caption{ Stellar mass vs specific star formation rate (uncorrected for dust) for the same spiral galaxies (small dots) and ETGs (bigger symbols).
As in Figure~\ref{TFR}, the spirals are colour-coded according to their redshift, and the ETGs are blue if in the field, red if in a cluster/group, 
filled if undisturbed kinematics, open if disturbed, circle if elliptical and square if S0. No redshift or magnitude cuts were applied to the plotted sample. }
 \label{SFR_mass}
\end{figure}

Figure~\ref{SFR_mass} shows the sSFR as a function of stellar mass for the spirals and ETGs considered in this paper. 
Most of the ETGs have moderate-to-low sSFR in comparison with spirals. 
At the mean stellar mass (=5.63$\times10^{10}M_{\odot}$),  the sSFR of high-redshift spirals is $\sim 4$ times higher than that of ETGs. 
And if we take the intermediate-redshift galaxies, the difference is still a factor of 2. 
%
Most strikingly, our cluster emission-line ETGs are typically much less massive on average than field emission-line ETGs, and the  field emission-line ETGs have the lowest sSFRs. The low sSFRs found in the ETGs, together with their red colours, is consistent with their location in the TFR.

Our results are in line with \citet{Verdugo2008}, who found that cluster ellipticals at $z=0.25$ have lower sSFRs than cluster spirals. Intriguingly, they also detect emission lines in the spectra of some of their red-sequence cluster galaxies, for which they, however, lacked morphological information. 
They find that the sSFRs of these emission-line red-sequence galaxies are  between those of ellipticals and spirals. 
We can speculate that their objects might be lower-redshift analogues of our emission-line ETGs. 

Moreover, our results agree with the SAURON study of ETGs in the local Universe \citep{Shapiro2010}, who found that nearly all star-forming ETGs fall below the main sSFR-stellar mass sequence of actively star-forming galaxies, but nonetheless have more star-formation activity than the quiescent population.

\section{Discussion}
\label{sec:discusion}

Despite the difficulties imposed by the high-redshift nature of our sample, we have found clear evidence for the existence of ETGs with extended (sometimes rotating) ionized gas. Our findings are in agreement with low-redshift studies of elliptical and S0 galaxies \citep[e.g.][]{Trager2000,Cappellari2011b}, making  this sample the first higher-redshift analogue of the population of ETGs with complex ISM seen in the local Universe. 

Two main questions arise from the finding of ETGs with ionized gas: where did the gas come from, and what ionizes it?

Considering the latter question, the spectral properties of the galaxies under study suggest that star formation is the most likely explanation for the ionization observed. However, we cannot completely rule out possible contributions from post-AGB stars or AGN activity (see Section~\ref{subsec:ionization}). 

As for the origin of the gas in ETGs,  there are two  possibilities: (a) acquisition from an external source, and (b) internal production via mass return into the ISM from evolved stars. These two hypothesis have been debated in the literature for years \citep[e.g.][]{ FaberGallagher1976, Sarzi2006}. We discuss our results in the context of each scenario in the following.

\textbf{(a)} 
In the first scenario, field elliptical and S0s with extended emission acquired their gas after they acquired their morphology, from an external source. This is possible because they do not inhabit the hostile cluster environment that could have prevented gas accretion. Instead, in the field, galaxy-galaxy mergers are more likely to occur. 

If we assume star formation to be the main source of ionization of the gas in these ETGs, the small amount of star formation that we find could thus be interpreted as a ``frosting'' on top of the old stellar population that dominates these galaxies. This frosting could be caused by minor gas-rich mergers, as the morphologies of these ETGs do not appear to be perturbed. This scenario is in line with CDM models of structure growth predicting minor mergers and gas accretion still at lower redshifts.

\textbf{(b)} 
However, the ``observed star formation" can also be explained by gas returned from previously formed stars, which is expected from stellar evolution models. 
We estimate that $\sim20-40\%$ of the total stellar mass is returned as gas by the evolved stellar populations\footnote{The exact value depends on the initial mass function assumed}. If this gas is retained by the galaxy, it can form a significant amount of stars, and drive substantial sSFR. For instance, a galaxy with a stellar mass of $10^{11} M_{\odot}$ will have $\sim 0.3 \times 10^{11} M_{\odot}$ of gas returned. If all this gas were then  converted into stars and the star formation episode lasts for 5 Gyr, the SFR would be $6M_{\odot}/yr$ (equivalent to sSFR$=6\times10^{-11} yr^{-1}$), which is higher than the values found for $10^{11} M_{\odot}$ galaxies (see Figure~\ref{SFR_mass}). 
If we assume a more realistic case in which only a few percent of the gas is efficiently transformed into stars, we obtain the sSFRs observed in our sample. 
In other words, gas recycling alone can explain the levels of star formation found in the ETGs considered. 
This is expected to be possible only in the field, because cluster environment can remove gas from galaxies by means of for example ram-pressure stripping.  
This scenario is consistent with (1) field galaxies retaining more gas, thus being more likely to form stars and (2) undisturbed field ETGs (solid blue symbols in Figure~\ref{SFR_mass}) having higher sSFR than disturbed ETGs (open blue symbols), which have presumably experienced gas loss. If the gas has been accreted from an external source, the star formation of the kinematically disturbed galaxies would be expected to have been enhanced. 
However, some models predict a suppression of star formation, so we cannot completely rule out an external origin. 

Our conclusions are also in agreement with the study of Rudnick et al. (in preparation), who concludes from a mass-selected sample of EDisCS old galaxies, that the ionized gas observed in these systems is potentially stripped by ram-pressure in clusters. 

Future observations with current and future state-of-the art facilities will unveil the origin of the gas in ETGs at increasingly high redshift. In particular, resolved 2D kinematics for both the stellar and gaseous components in these high-redshift objects can place strong constraints in these galaxies' history.

\section{Summary and Conclusions}
\label{sec:conclusions}

We have studied the properties of the gas and the stars in a sample of 24 emission-line E/S0 galaxies from EDisCS in different environments at $0<z<1$. Our main findings are as follows:

\begin{itemize}

\item ETGs with extended emission represent a significant fraction ($\geq12\%$) of the emission-line galaxy population at $z<1$.

\item We explore the possible ionizing sources of the gas in the ETGs at $z<1$, such as AGN activity, star formation, and post-AGB stars, and favour star-formation as the most likely cause of the observed emission.

\item Many of the emission-line ETGs have colours consistent with  the old stellar populations found in the red sequence. However, 47\% have colours that are bluer than 0.3 (and up to 1) mag from the model predictions for passive evolution after a burst of star formation. This implies that there is a small amount of on-going  or recent ($\lesssim 1$Gyr) star formation, involving as little as $\sim5$\% of the galaxy's stellar mass.

\item These emission-line ETGs are only found in the field and  infall regions of clusters. 
 This has different implications, depending on the origin of the gas:
 \begin{enumerate}
  \item If the gas was acquired via interactions with other galaxies, 
this gas will be longer-lived outside the harsh cluster environment. This
result is in agreement with the idea that ram pressure strips the gas in galaxies at the centres of clusters. 
  \item If instead the gas has an external origin, it is reasonable to expect to detect it in galaxies inhabiting the field or cluster outskirts, as they are the most likely places to host galaxy-galaxy interactions. 
 \end{enumerate}

\item Some (20\%) of the ETGs with emission show signs of moderate interaction in their stellar light, but most are unperturbed ellipticals or S0s. This indicates that, if the gas has an external origin, it was acquired after the galaxy had already achieved its current morphology and in a process gentle enough not significantly affect that morphology.

\item We analyse the TFR of the emission-line ETGs and compare them with EDisCS spirals, taking a matched range of redshifts, $M_B < -20$, and excluding galaxies with large velocity uncertainties. We find that EDisCS emission-line ETGs are $\sim$1.7 mag fainter (in the rest-frame $B$-band) than spirals at a fixed rotational velocity. This result is consistent with local TFR studies, suggesting that ETGs have lower mass-to-light ratios than spirals since their light is dominated by older stellar populations.

\item Assuming star formation to be the main source of ionizing radiation, we find that, at fixed stellar mass, cluster emission-line ETGs have lower sSFRs than  spirals, and field emission-line ETGs have, in general, the lowest sSFRs of the entire emission-line galaxy population in EDisCS. 
We emphasize, however, that our emission-line ETGs have higher SFRs  than typical ETGs. 
Our results agree with low-redshift studies such as that of \citet{Verdugo2008} and \citet{Shapiro2010}. 

\end{itemize}

The analysis presented here constitutes an important step forward towards the understanding of the evolution of the complex ISM in ETGs by significantly extending the look-back-time baseline of such studies which have until now concentrated almost exclusively on low-redshift samples. Observations with current and future state-of-the art facilities will reveal further details about the ISM in ETGs at increasingly higher-redshifts.

\section*{Acknowledgements}
We would like to thank the anonymous referee for helping improve this paper significantly. 
The work here presented is based on observations collected at the European Southern Observatory, Chile, as part of programme 166.A-0162.
YLJ gratefully acknowledges support by FONDECYT grant N. 3130476. 
RD acknowledges the support provided by the BASAL Center for Astrophysics and Associated Technologies (CATA), and by FONDECYT grant N. 1130528.



\appendix
\section{Data table}
Table~\ref{all_table} lists the main properties of all the ETGs described in Section \ref{sec:el_etg_sample}.

\begin{table*}
 \begin{center}
\caption{The emission-line ETG sample considered in this paper is listed in the first 24 rows. At the bottom, the galaxies with artificial emission (i.e., galaxies whose apparent emission is an observational artefact, see section~\ref{subsec:kinematics}), and the galaxies that were misclassified as ETGs are also listed for reference. The columns are: EDisCS name, environment (``c" for cluster, ``f" for field and ``g" for group), redshift, absolute $B$-band magnitude, rotational velocity, flag for emission-line kinematics (``good" is undisturbed, ``dist" is disturbed, and ``bad" is artificial emission), Morphology, flag for morphological disturbance (``good" is undisturbed, ``dist" is disturbed). When an entry has ``--" it means there is no data available. In addition, for ETGs with real emission (first 24 rows), we include 3  extra columns indicating their effective radii ($r_{\rm eff}$, computed from the single-S\'ersic fits), the ration between the spatial extent of the emission \citep[$r_{\rm extent}$, see][for details]{Jaffe2011b} and $r_{\rm eff}$, and the equivalent width of the [OII] doublet. }
\begin{scriptsize}
\label{all_table}
\begin{tabular}{lcccccccccc}
\hline 
Object ID 	&  envi-.&   $z$	&  $M_{B}$	&  $\log V_{\rm rot}$	 	& kinem.&   Morph. 	&  morph.  & $r_{\rm eff}$  &  $r_{\rm extent}$/$r_{\rm eff}$  & EW[OII] \\
(EDCSNJ*) 	&  ronment &   		&  (mag)	&  (km s$^{-1}$)			  &  dist.&   	&  dist. & (kpc)  &    & (\AA)  \\
\hline
\textbf{Disc kinematics} \\
1103485-1247452	& f	& 0.7668	& $-$20.50	&1.19$^{+0.54}_{-\rm{INDEF}}$ & good	&       E	&dist$^{\dagger \dagger}$	& -- 	& -- 	& 23.511	\\
1138115-1135008	& f	& 0.1857	& $-$16.90	&2.02$^{+0.04}_{-0.05}$ 		 & good	&       E	&good& 0.81	& 3.35	& -- \\
1138170-1131411	& f	& 0.2605	& $-$19.23	&1.87$^{+0.08}_{-0.10}$ 		 & good	&       E	&dist& 1.05	& 3.81	& --		\\
1227507-1139384$^{\dagger}$	& f	& 0.8725	& $-$22.66	&1.58$^{+0.39}_{-\rm{INDEF}}$ & good	&       E	&good& 4.48	& 1.80	 & 9.45	\\
1227577-1137211	& f	& 0.5451	& $-$20.83	&2.46$^{+0.03}_{-0.03}$ 		 & good	&       E	&good& 3.57	& 2.93	& 9.869	\\
1228026-1139163$^{\dagger}$	& f	& 0.3431	& $-$17.46	&1.05$^{+0.30}_{-\rm{INDEF}}$ & good	&       E	&good& 0.63	& 10.69	& --	\\
1354055-1234136	& f	& 0.5142	& $-$21.47	&2.39$^{+0.07}_{-0.13}$ 		& good	&       E	&dist& 3.41	& 0.51	& 3.998	\\
1354073-1233336	& c	& 0.7670	& $-$19.95	&2.17$^{+0.02}_{-0.04}$ 		& good	&       E	&good& 1.7	& 4.09	& 29.349	\\
1354074-1233206	& f	& 0.8177	& $-$20.35	&2.22$^{+0.01}_{-0.01}$ 		& good	&       E	&good& 1.59	& 4.10	& 60.993 	\\
1037495-1246452$^{\dagger}$	& f	& 0.5327	& $-$22.05	&2.57$^{+0.03}_{-0.05}$ 		& good	&       S0	&dist&	6.18 & 0.95	&17.178	\\
1037553-1246380	& g	& 0.5768	& $-$19.83	&1.58$^{+0.17}_{-0.27}$ 		& good	&        S0	&good&1.64 	&2.92	&30.476	\\
1054207-1148130$^{\dagger}$	& c	& 0.6996	& $-$19.98	&1.00$^{+0.66}_{-}$	& good	&        S0	&good&2.29	&1.28	&	27.794	\\
1138112-1135117$^{\dagger}$	& c	& 0.4842	& $-$18.36	&0.88$^{+0.53}_{-\rm{INDEF}}$ & good	&        S0	&good& 1.62	& 1.26	& 19.429	\\
1232274-1251372	& f	& 0.1467	& $-$15.28	&1.52$^{+0.12}_{-0.13}$ 		& good	&        S0	&good& 1.26	& 2.27	& -- 	\\

\hline 
\textbf{Disturbed kinematics} \\
1054525-1244189	& g	& 0.7283	& $-$22.34	&	--				& dist	&       E	&dist& 4.21	& 0.98	& 1.582	\\
1354144-1228536\_A& f	& 0.8245	& --	&	--				& --	&       E	&--&	 -- & --	 &	-- \\
1354144-1228536\_B	& f	& 0.8243	& --	&	--				& --	&       E	&--&	 -- & 	-- & --	\\
1354022-1234283	& c	& 0.7711	& $-$20.29	&	--				& dist	&       E	&good& 1.04	& 3.93	& 62.848		\\
1040356-1156026	& c	& 0.7081	& $-$21.69	&	--				& dist	&       E	&good& 2.66	& 10.68	& 2.908		\\
1040415-1156207	& f	& 0.6240	& $-$19.32	&	--				& dist	&        S0	&good& 1.97	& 0.48	& 9.088	\\
1054356-1245264	& c	& 0.7493	& $-$21.89	&	--				& dist	&        S0	&dist& 2.49	& 1.50	& 5.807	\\
1138069-1136160	& c	& 0.4520	& $-$18.62	&	--				& dist	&        S0	&good& 2.08	& 1.278	& 15.312	\\
1216446-1202358	& f	& 0.6698	& $-$20.75	&	--				& dist	&        S0	&good& 1.47	& 3.33	& 8.773	\\
1354107-1231236	& f	& 0.6183	& $-$21.96	&	--				& dist	&        S0	&dist& 1.63	& 1.58	& 2.446	\\
\hline
\hline 
\textbf{Artificial emission} \\
1054339-1147352	& f	& 0.8608	& $-$21.97	&	--				& bad	&       E	&good&	&	&	\\
1103458-1243353	& f	& 0.4275	& $-$20.12	&	--				& bad	&       E	&good&	&	&	\\
1138116-1134448	& c	& 0.4571	& $-$20.72	&	--				& bad	&       E	&dist&	&	&	\\
1138204-1131417	& f	& 0.9090	& $-$21.75	&	--				& bad	&       E	&good&	&	&	\\
1216527-1202553	& f	& 0.8263	& $-$21.45	&	--				& bad	&       E	&good&	&	&	\\
1216541-1157559	& f	& 0.8748	& $-$21.75	&	--				& bad	&       E	&good&	&	&	\\
1216548-1158039	& f	& 0.9827	& $-$21.46	&	--				& bad	&       E	&--	&	&	&   \\
1227589-1135135	& c	& 0.6375	& $-$22.91	&	--				& bad	&       E	&good&	&	&	\\ 
1354139-1229474	& f	& 0.6865	& $-$22.05	&	--				& bad	&       E	&good&	&	&	\\
1354185-1234431	& f	& 0.9092	& --			&	--				& bad	&       E	&--&		&	&	\\
1037552-1246368	& c	& 0.4245	& $-$20.93	&	--				& bad	&       S0	&dist&	&	&   \\    
1040476-1158184	& f	& 0.6171	& $-$20.85	&	--				& bad	&        S0	&good&	&	&	\\
1054436-1244202 & c & 0.7463& $-$21.49   &   --            	& bad    &       S0  &good&	&	&	\\
1103418-1244344	& f	& 0.3539	& $-$19.94	&	--				& bad	&        S0	&dist&	&	&	\\
1103430-1245370	& f	& 0.6584	& $-$21.64	&	--				& bad	&        S0	&dist&	&	&	\\
1227552-1137559	& f	& 0.4893	& $-$21.22	&	--				& bad	&        S0	&dist&	&	&	\\
1232288-1250490	& c	& 0.5470	& $-$22.30	&	--				& bad	&        S0	&good&	&	&	\\

\hline 
\textbf{Misclassified ETGs} \\
1138086-1131416	& f	& 0.5039	& $-$19.03	&1.55$^{+0.17}_{-0.27}$ 		& good	&       Irr/Disc	&dist&	&	& \\
1354049-1234087	& f	& 0.6617	& $-$20.89	&2.48$^{+0.03}_{-0.03}$ 		& good	&       Spiral	&good&	&	& \\
1354176-1232261	& f	& 0.4779	& $-$20.50	&2.54$^{+0.01}_{-0.01}$ 		& good	&       Spiral	&dist&	&	& \\

\hline

\end{tabular} 
\\
$^{\dagger}$ These emission-line galaxies have velocities and velocity errors that makes them \\
 consistent with no rotation and are thus not plotted in the TFR of Figure~\ref{TFR}. \\
 $^{\dagger \dagger}$ This galaxy is at the edge of the HST image and hence we did not attempt a single-Sersic fit to it.\\
\end{scriptsize}
\end{center}
\end{table*}

\clearpage
\newpage

\section{Morphologies and kinematics}
\begin{figure*}
\begin{center}
  \includegraphics[width=0.9\textwidth]{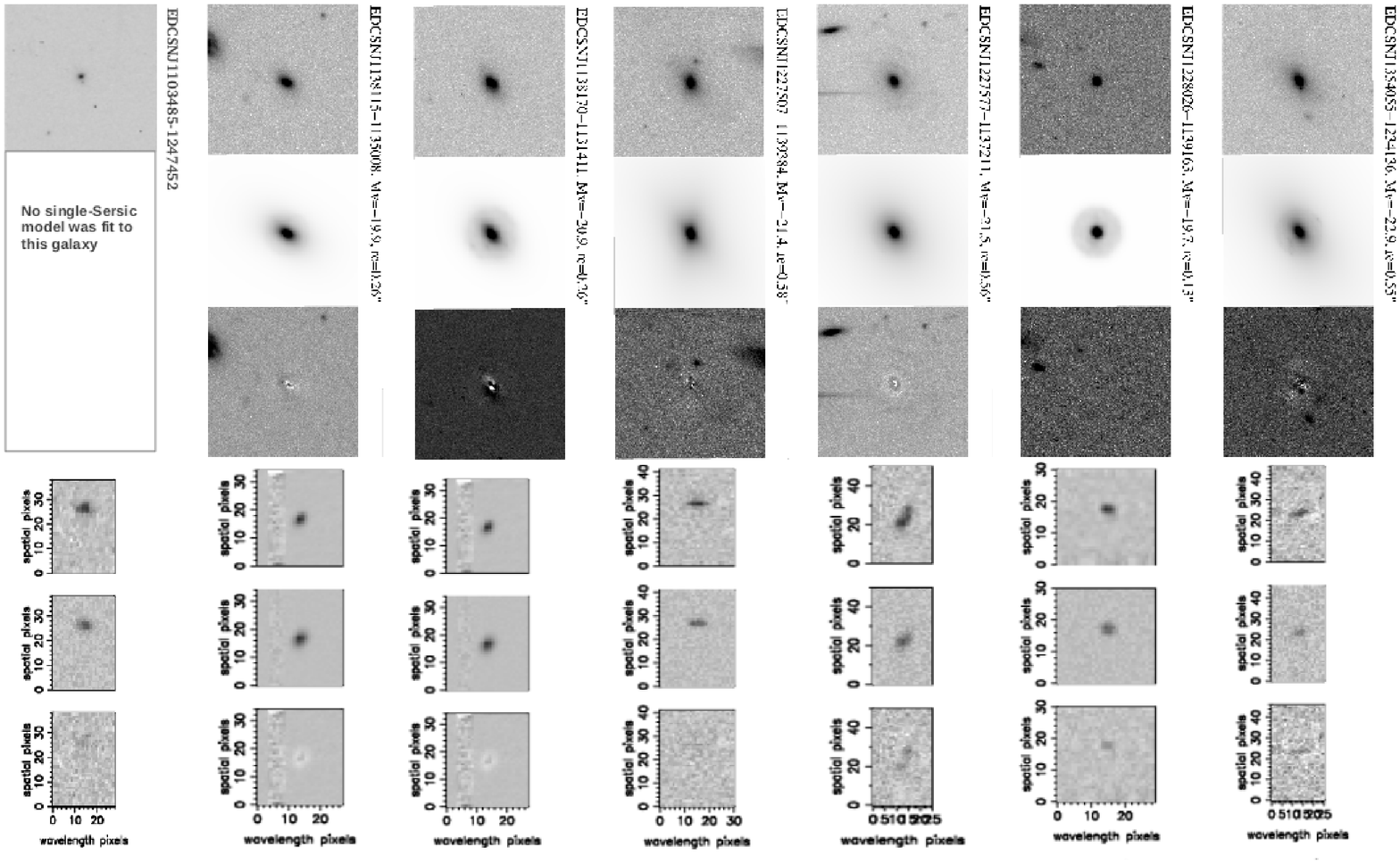}
  
  \includegraphics[width=0.9\textwidth]{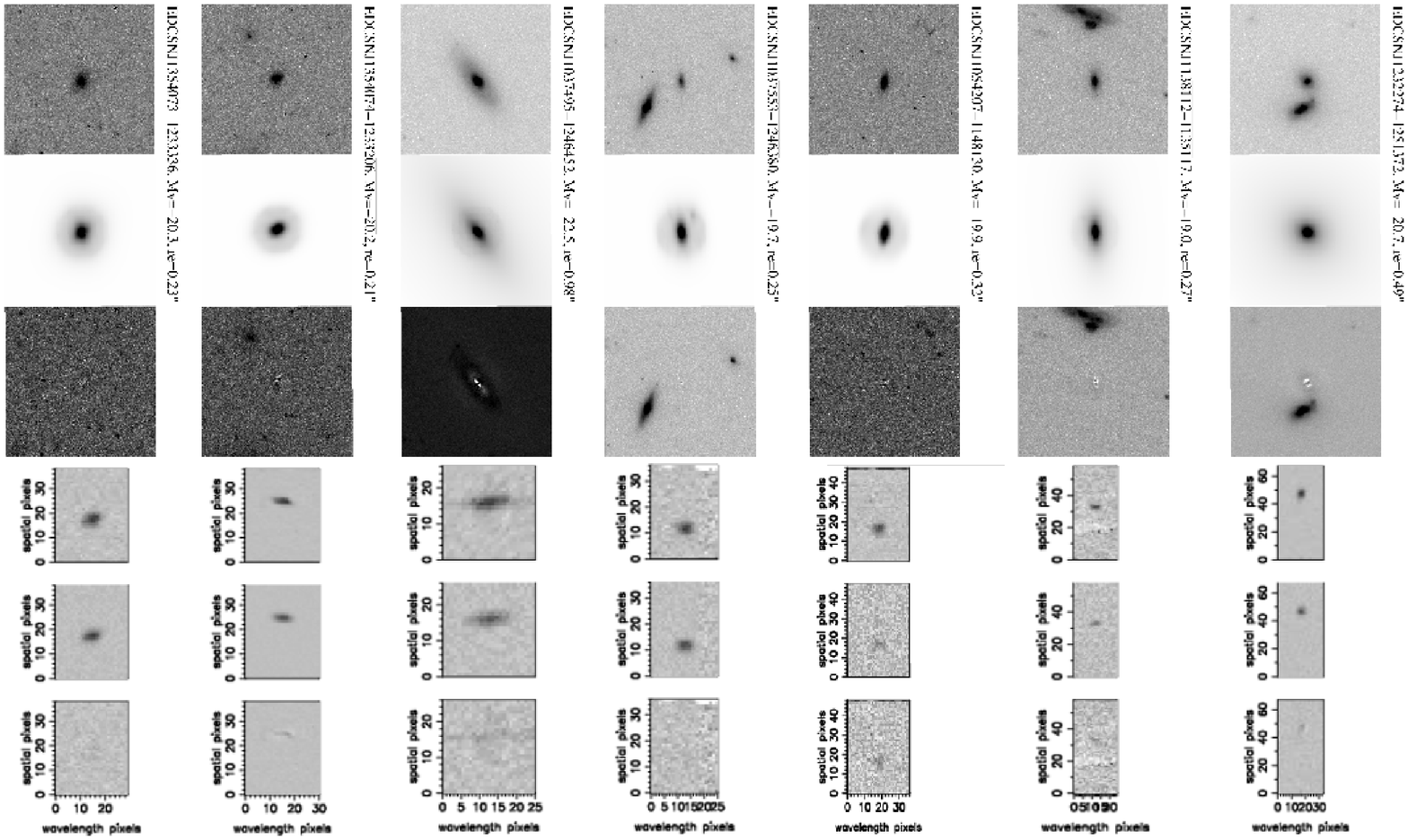}
\end{center} 
\caption{The 14 E/S0 in our sample showing signs of rotation in their gas emission (i.e. undisturbed kinematics) in 2 rows of 6 panels each. For each galaxy, the upper panel shows a 4"$\times$4" HST image of the galaxy, the next panel the single-Sersic fit, and the one below the residual image. The following 3 lower panels show the most prominent emission line, the 2D model, and the residual respectively. 
The emission lines shown here are mostly the [OII]3727$\rm \AA$ doublet. When [OII]3727$\rm \AA$ was not available, the [OIII]5007$\rm \AA$ line is shown instead (these cases can be identified by an ``--" in column 11 of Table~\ref{all_table}). 
Note that for the galaxy EDCSNJ1103485-1247452 it was not possible to perform a good Single-Sersic fit to it.}
 \label{HST_mosaic_ETG_rot}
\end{figure*}

\begin{figure*}
\begin{center}
  \includegraphics[width=0.78\textwidth]{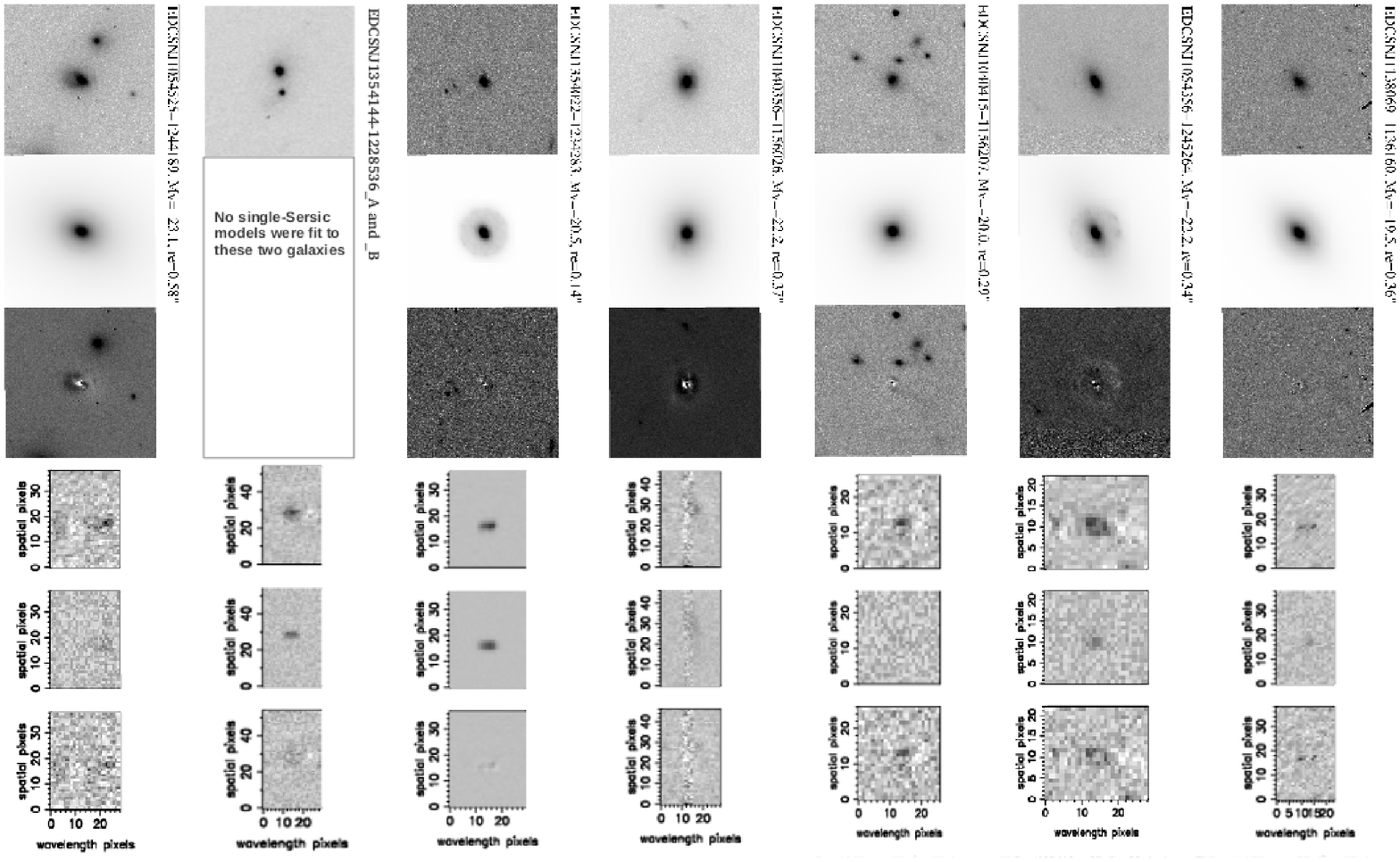}
   \includegraphics[width=0.205\textwidth]{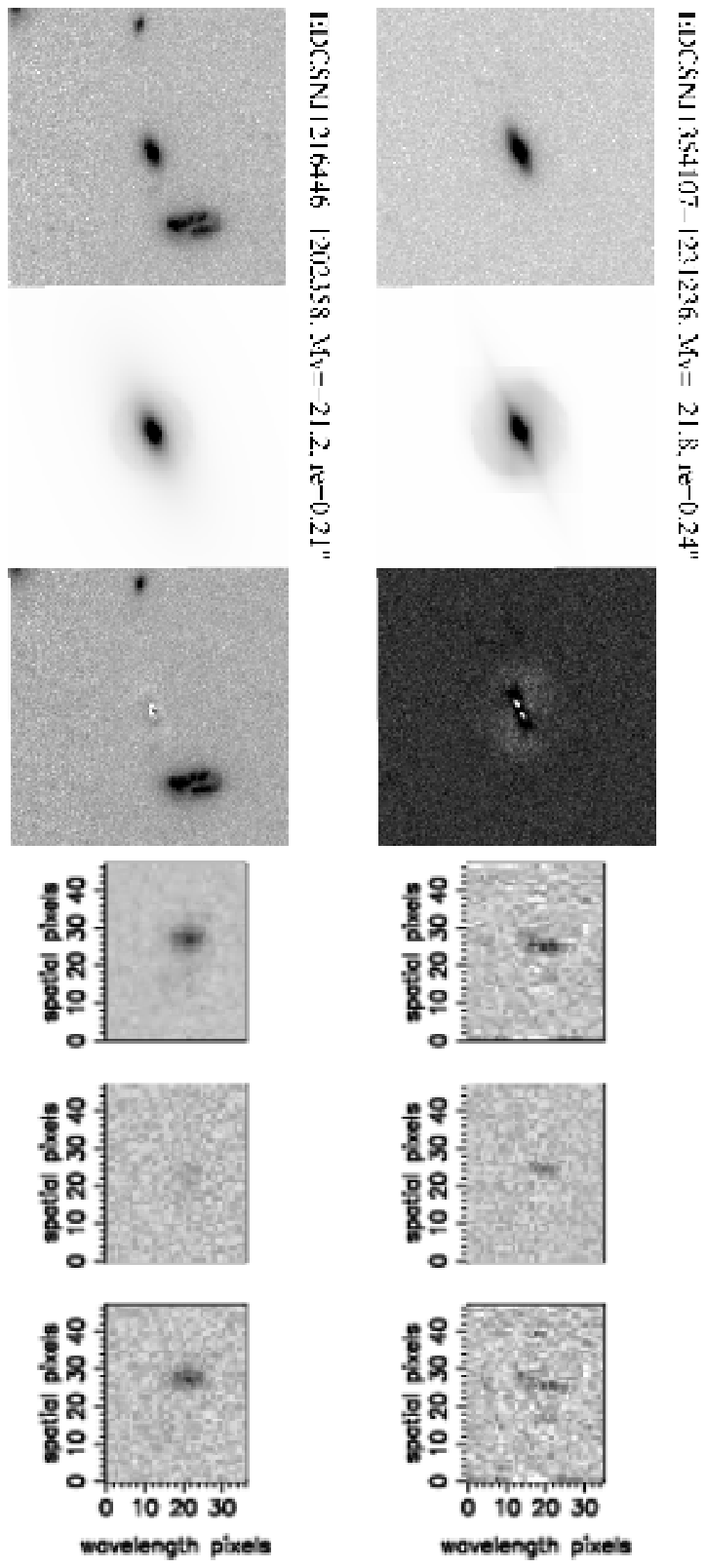}
\end{center} 
\caption{The 10 E/S0 in our sample showing extended emission but with disturbed gas kinematics. The panels are as in Figure~\ref{HST_mosaic_ETG_rot}. Note that the pair EDCSNJ1345144-1228536\_A and \_B are displayed together due to their vicinity.}
 \label{HST_mosaic_ETG_dist}
\end{figure*}

In the following we show the postage-stamps of the HST images, single-sersic models, emission-lines in the optical spectra and the respective 2D emission-line fit for the 24 galaxies with E/S0 morphology and real emission spectra. As in Table~\ref{all_table}, we separate the sample into kinematically disturbed and undisturbed galaxies (Figures~\ref{HST_mosaic_ETG_dist}~and~\ref{HST_mosaic_ETG_rot} respectively).

\end{document}